\documentclass[a4paper,accepted=2026-05-20]{quantumarticle}
\pdfoutput=1
\usepackage[utf8]{inputenc}
\usepackage[english]{babel}
\usepackage[T1]{fontenc}
\usepackage{amsmath}
\usepackage{tikz}
\usepackage{lipsum}
\usepackage{amsmath}
\usepackage{amsfonts}
\usepackage{amssymb}
\usepackage{dsfont}
\usepackage{physics}
\usepackage{setspace}
\usepackage{dsfont}
\usepackage{xcolor}
\usepackage{bbold}
\usepackage[numbers]{natbib}
\usepackage{printlen}
\usepackage{url}
\usepackage[colorlinks = true, linkcolor = blue, urlcolor  = blue, citecolor = blue, anchorcolor = blue]{hyperref}
\usepackage{pgfplots}
\usetikzlibrary{automata,arrows,positioning,calc}
\usepackage{tikz}
\usepgfplotslibrary{fillbetween}
\usetikzlibrary{arrows.meta, shadows, fadings,shapes.arrows,positioning}
\usepackage[left=3cm,right=3cm,top=3cm,bottom=3cm]{geometry}

\newcommand{\Smat}{\mathcal{Z}}
\newcommand{\genpsi}{\Psi}

\usetikzlibrary{calc}
\usetikzlibrary{arrows.meta}
\pgfplotsset{compat=newest}

\newcommand{\e}{\mathrm{e}}
\renewcommand{\i}{\mathrm{i}}

\begin{document}
\title{Squeezing Enhancement in Lossy Multi-Path Atom Interferometers}
\author{Julian Günther}
\affiliation{Institute for Theoretical Physics, Leibniz Universität Hannover, Appelstraße 2, 30167 Hannover, Germany}
\affiliation{Institute of Quantum Optics, Leibniz Universität Hannover, Welfengarten 1, 30167 Hannover, Germany}
\author{Jan-Niclas Kirsten-Siemß}
\affiliation{Institute of Quantum Optics, Leibniz Universität Hannover, Welfengarten 1, 30167 Hannover, Germany}
\author{Naceur Gaaloul}
\affiliation{Institute of Quantum Optics, Leibniz Universität Hannover, Welfengarten 1, 30167 Hannover, Germany}
\author{Klemens Hammerer}
\affiliation{Institut für Theoretische Physik, Universität Innsbruck, 6020 Innsbruck, Austria}
\affiliation{Institute for Quantum Optics and Quantum Information, Austrian Academy of Sciences, 6020 Innsbruck, Austria}
\affiliation{Institute for Theoretical Physics, Leibniz Universität Hannover, Appelstraße 2, 30167 Hannover, Germany}
\email{Klemens.Hammerer@itp.uni-hannover.de}

\begin{abstract}
This paper explores the sensitivity gains afforded by spin-squeezed states in atom interferometry, in particular using Bragg diffraction. We introduce a generalised input-output formalism that accurately describes realistic, non-unitary interferometers, including losses due to velocity selectivity and scattering into undesired momentum states. This formalism is applied to evaluate the performance of one-axis twisted spin-squeezed states in improving phase sensitivity. Our results show that by carefully optimising the parameters of the Bragg beam splitters and controlling the degree of squeezing, it is possible to improve the sensitivity of the interferometer by several dB with respect to the standard quantum limit despite realistic levels of losses in light pulse operations. However, the analysis also highlights the challenges associated with achieving these improvements in practice, most notably the impact of finite temperature on the benefits of entanglement. The results suggest ways of optimising interferometric setups to exploit quantum entanglement under realistic conditions, thereby contributing to advances in precision metrology with atom interferometers.
\end{abstract}

\maketitle

\section{Introduction}

The enhancement of interferometry using entangled states, particularly squeezed states has become a routine practice in gravitational wave observatories~\cite{LIGO_Ganapathy2023}, and is a highly researched area in frequency metrology~\cite{Colombo2022} and atom interferometry~\cite{Szigeti2021,Szigeti,Corgier2021,Corgier2021b,Salvi2018}. In the domain of atom interferometry, squeezing-enhanced setups, the focus of the current paper, have been  realised in~\cite{Anders2021,Malia2022,Greve2022}, and recently applied successfully for gravimetry in~\cite{Cassens2024}. This holds the perspective for improving atom interferometric measurements of the fine-structure constant~\cite{Morel2020,Parker_2018}, gradiometry~\cite{Asenbaum2017,Chiow2017,Biedermann2015,McGuirk2002}, tests of general relativity~\cite{Dimopoulos2007,Dimopoulos2008,Ufrecht2020,Asenbaum2020,Werner2024}, and the detection of gravitational waves~\cite{Abend2024,Abe_2021,Badurina_2020,Canuel_2020,Tino_2019,Canuel_2018,Hogan2016,Hogan_2011,Dimopoulos2008b}.

In general, entanglement-enhanced interferometry with squeezed states in principle holds the potential to achieve a \(1/\sqrt{N}\) improvement over the standard quantum limit in interferometry~\cite{RevModPhys.90.035005}. However, the impact of noise and losses under realistic conditions often makes realizing this improvement challenging or even impossible~\cite{ElusiveHeisenberg,Escher2011}. Consequently, the optimal enhancement must be determined through a detailed, case-by-case analysis, considering the specific characteristics of the given interferometric platform.

In this study, we consider atom interferometers based on Bragg diffraction~\cite{Giltner1995,Martin1988}, which currently offer the largest metrological scale factor~\cite{Rodzinka2024}. Bragg diffraction enables highly efficient, yet inherently lossy, light-pulse operations for atom interferometry~\cite{Mueller2008,Giese2013,Siem__2020,AdaptedMirror}. The dominant loss mechanisms, such as Doppler detunings leading to velocity selectivity and diffraction into undesired momentum states, must be carefully analysed~\cite{Szigeti_2012} to evaluate the potential improvements in phase sensitivity from using squeezed states.

Evaluating the squeezing enhancement while accounting for Doppler effects and the multi-path, multi-port nature inherent to Bragg diffraction is a non-trivial task. It requires tracking the propagation of a correlated \(N\)-particle wave function through the interferometer. To address this, we develop a general formalism for monitoring the first and second moments using polarization vectors and covariance matrices of the relevant pseudo-spins that describe the initial state of the atoms and the detected output ports of the interferometer. 

This formalism is applied to a case study involving a Mach-Zehnder interferometer (MZI) with a squeezed input state. We focus specifically on the scheme proposed by Szigeti et al.~\cite{Szigeti}, which utilizes s-wave scattering to generate squeezed states via one-axis-twisting. We derive the optimal light-pulse parameters to maximize squeezing enhancement, considering finite momentum width (Doppler effects) and undesired diffraction orders. Our findings reveal that while squeezing enhancement can be significant, the details of the input states—such as the levels of squeezing and momentum width—and the characteristics of the interferometer's light-pulse operations crucially determine the efficacy of entanglement enhancement.

The paper is organised into two main sections. In Section 2, we develop the generalised input-output formalism for the treatment of entangled input states in lossy atom interferometers. Section 3 focuses on the specific example of a MZI based on Bragg diffraction, where we determine the optimal light-pulse operations for maximising squeezing enhancement.

\section{Theoretical Framework}
\begin{figure}[t]
\includegraphics[width=\columnwidth]{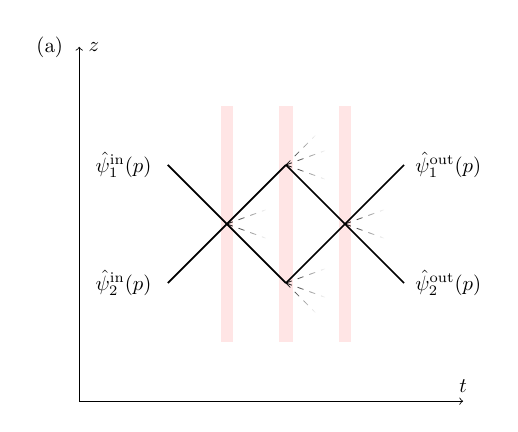}
\includegraphics[width=\columnwidth]{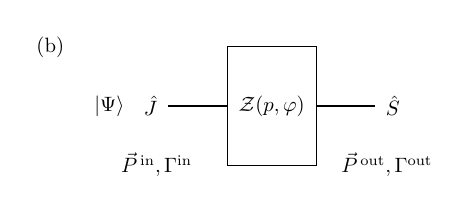}
\caption{\textbf{Schematic of an atom interferometer} (a) Space-time diagram of a MZI with two input and two output ports, each described by field operators $\hat{\psi}_i^{\mathrm{in(out)}}(p)$ in momentum space for port $i=1,2$. Light-pulse beam splitter and mirror operations based on Bragg diffraction are inherently lossy due to scattering into undesired diffraction orders (dashed lines). (b) The interferometer maps a pseudo-spin $\hat{J}$ in input state $\ket{\Psi}$ to an output pseudo-spin $\hat{S}$, whose first and second moments are described by the polarization vectors $\vec{P}^{\mathrm{in(out)}}$ and covariance matrices $\Gamma^{\mathrm{in(out)}}$. The transfer matrix of the interferometer in momentum space is $\Smat(p,\varphi)$, where $\varphi$ is the interferometer phase, and determines the input-output relations for the polarization vectors and covariance matrices.}
\label{fig: simplified IF}
\end{figure}

We consider atom interferometers based on Bragg diffraction of matter waves on timed light pulses, as schematically shown in Fig.~\ref{fig: simplified IF}a. The space-time diagram refers to a  MZI as an example, but the consideration developed in the following apply to any interferometer geometry with two input and two output channels. 
These channels (labelled by $i=1,2$) refer to two classes (bins) of momentum states centered around momenta differing by a multiple of $2\hbar k$, the momentum recoil experienced by atoms in a two-photon transition. For the concrete example shown in Fig.~\ref{fig: simplified IF}a, the momentum bins for channels $i=1,2$ would be e.g. $\mp n\hbar k$ if the Bragg beam splitter was an $n$-th order Bragg diffraction imparting $2n\hbar k$ of momentum. Accordingly, the width of these bins in momentum space will be $2\hbar k$, which we assume to be larger than the momentum width $\Delta p$ of the incoming atomic wave packet, whose corresponding wave function we denote by $\phi(p)$. In this limit of a sufficiently narrow wave packet, cf.~\cite{Kovachy_2015,Deppner_2021}, it is justified and convenient to associate with each of the relevant momentum bins a one-dimensional bosonic field, with corresponding momentum space creation and annihilation operators obeying 
\begin{align}
 \comm*{\hat{\psi}_i(p)}{\hat{\psi}^\dagger_j(p^\prime)}=\delta_{ij}\delta(p-p^\prime).   
\end{align}

The mirror and beam splitter operations based on Bragg diffraction will generate a net transfer matrix $\mathcal{Z}(p,\varphi)$ for the full interferometer sequence propagating each momentum component as
\begin{align}\label{eq:InOutPsi}
    \begin{pmatrix}
        \hat{\psi}^\mathrm{out}_1(p) \\ \hat{\psi}^\mathrm{out}_2(p)
    \end{pmatrix}=\mathcal{Z}(p,\varphi)    \begin{pmatrix}
        \hat{\psi}^\mathrm{in}_1(p) \\ \hat{\psi}^\mathrm{in}_2(p)
    \end{pmatrix}.
\end{align}
Here, $\varphi$ denotes the interferometer phase, which could arise due to gravity. With $\mathcal{Z}(p,\varphi)$, the output field operators will also depend on $\varphi$. This dependency will be suppressed in the following, and only explicitly written out when necessary. Formally, $ \mathcal{Z}(p,\varphi) $ is a $ 2 \times 2 $ matrix that would be unitary for an ideal, lossless interferometer and $ p $--independent when Doppler detunings in the Bragg diffraction processes are disregarded. Due to parasitic diffraction orders and interferometer paths, as indicated in Fig.~\ref{fig: simplified IF}a, the transfer matrix of an interferometer based on Bragg diffraction will ultimately never be unitary and may exhibit a highly non-trivial dependence on the interferometer phase \cite{AdaptedMirror}. The specific form of the transfer matrix can be derived for a given interferometer geometry, as will be done in Sec.~\ref{sec:MZI} for a MZI. In this section, we take $ \mathcal{Z}(p,\varphi) $ as being given and make statements about the general class of interferometers described by input-output relations as given in Eq.~\eqref{eq:InOutPsi}.

The signal measured in the output of the interferometer is the population difference in the two output channels $i=1,2$, and thus corresponds to the pseudospin operator $\hat{S}_3$
given by
\begin{align}\label{eq:Sz}
    \hat{S}_3 &= \frac{1}{2}\left(\hat{n}^\mathrm{out}_1 - \hat{n}^\mathrm{out}_2 \right), \\
    \hat{n}^\mathrm{out}_i &= \int\dd{p}\; \hat{\psi}^{\mathrm{out}\dagger}_i(p)\hat{\psi}^{\mathrm{out}}_i(p). 
\end{align}
It will be convenient to introduce also the remaining components of this pseudospin, 
\begin{align}\label{eq:Sdef}
    \hat{S}_\alpha=\frac{1}{2} \int\dd{p} \sum_{i,j=1}^2 \hat{\psi}^{\mathrm{out}\dagger}_i(p)[\sigma_\alpha]_{ij}\hat{\psi}^{\mathrm{out}}_j(p),
\end{align}
where $[\sigma_\alpha]_{ij}$ denotes the $ij$--component  of the Pauli matrix $\sigma_\alpha$ for $\alpha=1,2,3$. In order to account for the effects of losses and the ensuing non-unitarity of the interferometer transfer matrix $\mathcal{Z}(p,\varphi)$, it will be useful to also consider the case $\alpha=0$ for which $\sigma_0$ is the identity matrix and $\hat{S}_0=\frac{1}{2}\left(\hat{n}^\mathrm{out}_1 + \hat{n}^\mathrm{out}_2 \right)$ measures the total population in both output ports. 

From the measured signal $\expval*{\hat{S}_3}$, the phase $\varphi$ can be inferred with an uncertainty
\begin{align}\label{eq:sensitivity}
\Delta \varphi = \left.\frac{\Delta \hat{S}_3}{\left|{\frac{\partial \langle \hat{S}_3 \rangle}{\partial \varphi}}\right|}\right|_{\varphi=0},
\end{align}
as follows from simple error propagation. The ultimate goal is to explore how much this phase uncertainty $\Delta \varphi$ can be decreased (i.e., how much the phase sensitivity of the interferometer can be increased) by using entangled states of atoms as an input to a (generally lossy) interferometer. It is expected that, for a given non-unitary transfer matrix $\mathcal{Z}(p,\varphi)$, there will be optimal levels of entanglement. E.g. for an optical interferometer, this has been shown in~\cite{Demkowicz2013}. Conversely, the light-pulse parameters must be chosen optimally to tailor the transfer matrix to a given entangled input state in order to achieve optimal sensitivity enhancement.

These optimisations hinge on an evaluation of the phase uncertainty in Eq.~\eqref{eq:sensitivity} for a given interferometer transfer matrix and atomic input state $\ket{\Phi}$. This is a rather simple task if the interferometer is operated with independently prepared, uncorrelated atoms, in which case the $N$-body input state would be
\begin{align}\label{eq:tensorinputstate}
    \ket{\Phi}=\frac{1}{\sqrt{N!}}\qty[\hat{a}^\dagger_i]^N\ket{\mathrm{vac}},
\end{align}
and correspond to a simple tensor product. Here, 
\begin{align}\label{eq:mode}
    \hat{a}^\dagger_i = \int dp\; \phi(p) \hat{\psi}^\mathrm{in\dagger}_i(p)
\end{align}
and $\hat{a}_i$ denote the creation and annihilation operators corresponding to the specific mode determined by the input momentum-wave function $\phi(p)$. We assume, that the modes for every momentum bin are identical and suppress indices for the momentum wave function $\phi(p)$. The mode matching of the momentum bins is necessary for the squeezing process and has been investigated in ~\cite{Szigeti}. \newline
For a tensor product state as in Eq.~\eqref{eq:tensorinputstate}, evaluating the uncertainty $\Delta \varphi$ using Eqs.~\eqref{eq:InOutPsi} and \eqref{eq:Sz} reduces to a one-body problem and merely requires propagating each component of $\phi(p)$ through the interferometer using the transfer matrix $\mathcal{Z}(p,\varphi)$. The result of this would exhibit the standard quantum limit (SQL), $\Delta \varphi=1/\sqrt{N}$. In contrast, when the input state $\ket{\Phi}$ is an entangled, for example squeezed, state of $N$ atoms, momentum components of different atoms will be correlated. Accordingly, propagating the $N$-body wave function becomes much more complicated and in general quite cumbersome.

In order to cope with this difficulty, it is practical to introduce yet another pseudospin, this time referring to the input of the interferometer,
\begin{align}\label{eq:Jdef}
    \hat{J}_\alpha &= \frac{1}{2} \sum_{i,j=1}^2
        \hat{a}^\dagger_i
    [\sigma_\alpha]_{ij}
        \hat{a}_j,
\end{align}
where $\alpha = 0, 1, 2, 3$ as before. Due to the linearity of the interferometer, as expressed in Eq.~\eqref{eq:InOutPsi}, it is in fact sufficient to know the first and second moments of this pseudospin for an entangled input state $\ket{\Phi}$ in order to evaluate its phase sensitivity. These numbers are conveniently collected in the polarization 4-vector 
\begin{align}
\Vec{P}^\mathrm{in}_\alpha = \langle \hat{J}_\alpha \rangle    
\end{align}
and the $4\times 4$ covariance matrix of the input
\begin{align}
\Gamma^{\,\mathrm{in}}_{\alpha \beta} = \frac{1}{2} \langle \{\hat{J}_\alpha,\hat{J}_\beta\} \rangle-P^\mathrm{in}_\alpha P^\mathrm{in}_\beta.  
\end{align}
where all averages are understood with respect the given state $\ket{\Phi}$. Correspondingly, the polarization vector and covariance matrix of the output state are $\Vec{P}^{\,\mathrm{out}}_{\alpha} = \langle \hat{S}_\alpha \rangle$ and $\Gamma^{\,\mathrm{out}}_{\alpha \beta} = \frac{1}{2} \langle \{\hat{S}_\alpha,\hat{S}_\beta\} \rangle - P^\mathrm{out}_\alpha P^\mathrm{out}_\beta$. Knowing these is sufficient to determine the sensitivity, since $\Delta \hat{S}^2_3 = \Gamma^{\mathrm{out}}_{33}$ and $\partial_\varphi \expval*{\hat{S}_3} = \partial_\varphi  P^\mathrm{out}_3$. What is thus required, is a connection between input and output of polarization vectors and covariance matrices, as indicated in Fig.~\ref{fig: simplified IF}b.

\section{Results}
\subsection{Input-Output Relations for Polarization Vector and Covariance Matrix}

\begin{figure}
    \centering
    \includegraphics[width=\columnwidth]{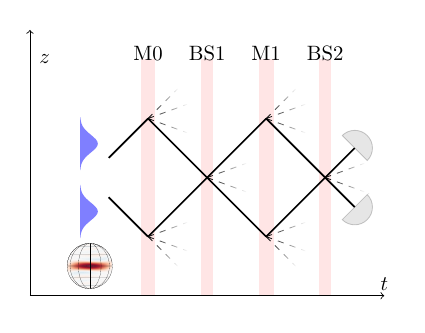}
    \caption{\textbf{MZI with squeezed input} In the scheme of Szigeti et al.~\cite{Szigeti}, a squeezed state (schemtically depicted on Bloch sphere) is generated by $s$-wave scattering during propagation through a first interferometer generated by a first beam splitter (not shown) and mirror pulse M0. In the main interferometer (generated by pulses BS1, M1, and BS2), the interferometer phase is picked up and detected in the output ports. Dashed lines indicate losses due to diffraction into undesired momentum states.}
    \label{fig:Szigeti IF}
\end{figure}

The main result of this section is an input-output relation that connects the polarization vectors and covariance matrices of the pseudospins $\hat{J}$ -- defined in Eq.~\eqref{eq:Jdef} for the input -- and $\hat{S}$ -- defined Eq.~\eqref{eq:Sdef} for the output. While the derivation involves some algebraic steps, which are detailed in  Appendix~\ref{App1}, the resulting input-output relation is surprisingly simple in its final form,
\begin{subequations}\label{eq:inout}
\begin{align}
    \Vec{P}^{\,\mathrm{out}} &= \mathcal{Q} \Vec{P}^{\,\mathrm{in}} \label{eq:Pinout}\\
    \Gamma^{\,\mathrm{out}} &= \mathcal{Q} \Gamma^{\,\mathrm{in}} \mathcal{Q}^T + \Gamma^\mathrm{noise}.\label{eq:Gammainout}
\end{align}
As is shown in the Appendix, momentum-dependent particle losses and parasitic interferometric signals are effectively accounted for in the noise matrix
\begin{align}\label{eq:Gammanoise}
\Gamma^\mathrm{noise}=\Lambda(\mathcal{Q}\Vec{P}^{\,\mathrm{in}}) - \mathcal{Q}\Lambda(\Vec{P}^{\,\mathrm{in}}) \mathcal{Q}^T.
\end{align}
\end{subequations}
The $4\times 4$ transfer-matrix $\mathcal{Q}$ is given componentwise by
\begin{align}\label{eq:Q}
\mathcal{Q}_{\alpha \beta} &= \int dp\: \abs{\phi(p)}^2  \frac{1}{2} \tr\left\{ \Smat^\dagger (p,\varphi) \sigma_\alpha \Smat (p,\varphi) \sigma_\beta 
\right\},
\end{align}
and depends on the transfer matrix $\mathcal{Z}(p,\varphi)$ of the interferometer and the input wave function $\phi(p)$. Below in Sec. \ref{sec:MZI}, we explicitly state $\mathcal{Z}(p,\varphi)$ for the example of a Mach-Zehnder interferometer in Fig.\ref{fig:Szigeti IF}. For a given $4$--vector $\Vec{P}$, the $4\times 4$ matrix $\Lambda(\Vec{P})$ used in the expression for the noise matrix is defined as
\begin{align}\label{eq:Lambda}
    \Lambda(\Vec{P}) &= \begin{pmatrix}
        P_0 & P_1 & P_2 & P_3 \\
        P_1 & P_0 & 0 & 0 \\
        P_2 & 0 & P_0 & 0 \\
        P_3 & 0 & 0 & P_0
    \end{pmatrix}.
\end{align}
We note that in general both $\mathcal{Q}$ and $\Gamma^\mathrm{noise}$ depend on the interferometric phase $\varphi$ via the transfer matrix $\mathcal{Z}(p,\varphi)$. The fact that the first and second moments of the input and output pseudospins $\hat{S}$ and $\hat{J}$ obey compact relations as in Eqs.~\eqref{eq:inout}, which are structurally identical to those for open-system dynamics of Gaussian states in bosonic modes, is the first main result of this work.

Some comments are in order regarding the input-output relations in Eqs.~\eqref{eq:inout}: Firstly, if the interferometer was ideal, that is lossless and free of Doppler effects, the transfer matrix would be unitary and independent of $p$. In this case, Eq.~\eqref{eq:Q} implies that $\mathcal{Q}$ takes on the block-diagonal form
\begin{align}
    \mathcal{Q}=\begin{pmatrix}
        1 & 0 \\
        0 & R(\varphi)
    \end{pmatrix}
\end{align}
where $R(\varphi)$ is a $3\times 3$ orthogonal matrix, which in turn ensues $\Gamma^\mathrm{noise}=0$, as follows from \eqref{eq:Gammanoise} and \eqref{eq:Lambda}. This recovers the limit of an ideal $SU(2)$ interferometer~\cite{Yurke1986} whose effect is to simply rotate polarization vectors and covariance matrices (as first and second order tensors, respectively). 

Secondly, Eqs.~\eqref{eq:inout} offer a noteworthy distinction between the aspects related to momentum width and Doppler effects, which are incorporated in $\mathcal{Q}$, and those associated with the type and strength of entanglement in the initial state, as reflected in $\vec{P}^\mathrm{in}$ and $\Gamma^\mathrm{in}$. The separation of these aspects is very useful for optimisations of quantum correlations for given interferometers and, vice versa, of pulse sequences for given entangled input states, as will be demonstrated in the next section.

Thirdly, the assumptions leading to Eqs.~\eqref{eq:inout} regarding the input state are minimal: only that the two input channels, $i=1,2$, are populated and that all atoms occupy the same mode defined by the momentum wave function $\phi(p)$, cf. Eq.~\eqref{eq:mode}. This assumption is based on expansion of the BEC's spatial mode in the interferometer that is needed to generate the squeezing in the first place~\cite{Szigeti}. No further assumptions are made about the specific quantum state. Therefore, these input-output relations can be applied to a wide range of input states of the form $\ket{\Phi}=f(a_1, a_1^\dagger, a_2, a_2^\dagger)\ket{\mathrm{vac}}$, where $f$ is an arbitrary function. This includes, for example, GHZ (NOON) states $[(a_1^\dagger)^N + (a_2^\dagger)^N]\ket{\mathrm{vac}}$ or one-axis-twisted states (OAT)~\cite{RevModPhys.90.035005},
\begin{align}\label{eq:OAT}
    \ket{T} = e^{-i\theta \hat{J}_1} 
    e^{-i\frac{\mu}{2}\hat{J}_3^2} \left(\frac{\hat{a}_{1}^\dagger + \hat{a}_{2}^\dagger}{\sqrt{2}}\right)^N \ket{\mathrm{vac}}
\end{align}
with $\mu\in\mathds{R}$ paramterizing the twisting strength. The angle $\theta$ controls a rotation of the state about the 1-axis, and will be addressed later on. The performance of these states will be investigated in the next section for the concrete example of a MZI.

\begin{figure}
    \centering
    \includegraphics[width=0.9\columnwidth]{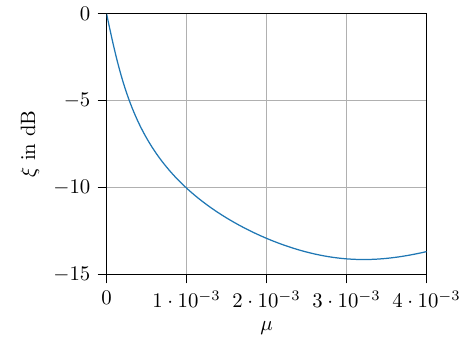}
    \caption{\textbf{Squeezing of OAT states} Wineland spin squeezing parameter $\xi$ as function of the twisting strength $\mu$ for $N=2\cdot 10^4$ particles.}\label{fig:xi_vs_mu}
\end{figure}

\subsection{Squeezing-enhanced high-order Bragg interferometry}\label{sec:MZI}

\subsubsection{Mach-Zehnder Interferometer with squeezed input states}

In the following, we will apply the general formalism developed in the previous section to the case of a MZI interferometer based on Bragg beam splitter and mirror operations with an input corresponding to a OAT state, as given in Eq.~\eqref{eq:OAT}. The specific geometry considered here is motivated by the scheme suggested, and worked out in great detail, by Szigeti et al.~\cite{Szigeti}. 

The setup of this proposal is shown schematically in Fig.~\ref{fig:Szigeti IF}. The main idea there is to run an auxiliary interferometer before the main metrological interferometer, and to keep the atomic density large enough during the propagation through the auxiliary interferometer in order to acquire a nonlinear phase in each arm due to $s$-wave scattering. It has been shown in~\cite{Szigeti} that this dynamics effectively generates an OAT state with a twisting parameter $\mu$ in Eq.~\eqref{eq:OAT}, which can be controlled by the propagation time in the auxiliary interferometer. Building on these results, we adopt an effective description of the scheme illustrated in Fig.~\ref{fig:Szigeti IF}, where the OAT state is assumed to be sequentially injected into both the auxiliary and main interferometers. Since our focus is on evaluating the impact of losses during the light-pulse operations, we consider the approximation that the entangled state is present from the start, rather than being gradually developed, as a worst-case scenario.

The main feature of the OAT state of Eq.~\eqref{eq:OAT} exploited here for achieving metrological enhancement is that it provides spin squeezing, that is, a reduction of spin projection noise along a particular direction transverse to its mean polarization along the 1-axis. The degree of squeezing is measured in terms of the Wineland squeezing parameter~\cite{RevModPhys.90.035005}
\begin{align}\label{eq:xi}
    \xi= \frac{\sqrt{N} (\Delta \hat{J}_2)}{\langle \hat{J}_1 \rangle }
\end{align}
We assume that the projection noise is reduced in the $2$-direction, which can be achieved without loss of generality by an appropriate choice of $\theta$ in Eq.~\eqref{eq:OAT}. In Fig.~\ref{fig:xi_vs_mu}, we show the dependence of the squeezing on the twisting parameter $\mu$. For sufficiently weak twisting, the squeezing parameter decreases monotonically, allowing us to express the input squeezing in terms of $\xi$ rather than the more abstract $\mu$. We will follow this convention in the subsequent discussion.

To determine the phase uncertainty using the input-output relations of Eqs.~\eqref{eq:inout}, we must follow these steps: \textit{(i)} determine the polarization vector $\Vec{P}^\mathrm{in}$ and the covariance matrix $\Gamma^\mathrm{in}$ for the given input state; \textit{(ii)} construct the transfer matrix $\mathcal{Z}(p,\varphi)$ of the interferometer under consideration; \textit{(iii)} calculate $\mathcal{Q}$ and $\Gamma^\mathrm{noise}$ to determine $\Vec{P}^\mathrm{out}$ and the covariance matrix $\Gamma^\mathrm{out}$, and finally derive the uncertainty $\Delta\varphi$ from this analysis.

Regarding \textit{(i)}, the input polarisation vector $\Vec{P}^\mathrm{in}$ and covariance matrix $\Gamma^\mathrm{in}$ for the OAT state have been worked out previously analytically in~\cite{Schulte2020} for a general number of atoms $N$ and squeezing parameter $\xi$ (or twisting strength $\mu$). Since the explicit expressions are somewhat bulky, we refrain from reproducing them here, and refer the reader to the given reference. 

Concerning \textit{(ii)}, the transfer matrix of the interferometer is constructed as a product of the transfer matrices from the mirror and beam splitter operations shown in Fig.~\ref{fig:Szigeti IF}, that is,
\begin{align}\label{eq:transferZ}
    \Smat(p,\varphi) = \Smat_{\mathrm{BS}}(p) G(\varphi) \Smat_{\mathrm{M}}(p) \Smat_{\mathrm{BS}}(p) \Smat_{\mathrm{M}}(p).
\end{align}
Here, we denote by $G(\varphi) = \textrm{diag}\qty(\exp(-i \varphi/2),\exp(i \varphi/2))$ the matrix accounting for free propagation and relative phase gain $\varphi$ among the two interferometer paths. The transfer matrices of individual mirror and beam splitter operations $\Smat_{\mathrm{M}}(p)$ and $\Smat_{\mathrm{BS}}(p)$ are in turn determined by numerically integrating the Schrödinger equation in momentum space. The procedure essentially follows the lines of~\cite{AdaptedMirror}, and is further detailed in Appendix~\ref{App2}. For the example considered here, we use third-order and fifth-order Bragg diffraction, imparting $6\hbar k$ and $ 10\hbar k$ of momentum recoil respectively, driven by Gaussian pulses characterised by a peak Rabi frequency $\Omega_0$ and duration $\tau$. A previous publication~\cite{Siem__2020} developed an analytic pulse-area relation that links the parameter pair $(\Omega_0, \tau)$ to achieve either a mirror or a beam splitter. For mirror operations $\Smat_{\mathrm{M}}(p)$, we fix $(\Omega_0, \tau)$ such as to realise the adapted mirrors introduced in~\cite{AdaptedMirror}, and realised recently in \cite{pfeiffer2024dichroicmirrorpulsesoptimized}. These mirror pulses are designed specifically to reflect the main interferometer paths back, while being maximally transparent to dominant parasitic paths, thereby avoiding closing parasitic interferometers. We note that the fact that the overall transfer matrix $\Smat(p,\varphi)$ can be understood as a product of the $2\times 2$ transfer matrices of individual pulses depends crucially on the fact that there are no parasitic interferometer paths. For the beam splitter pulses, $\Smat_{\mathrm{BS}}(p)$ will depend on the peak Rabi frequency $\Omega_0$, which can be tuned to achieve highest sensitivity, and will be used as a control parameter in the following. 

For step \textit{(iii)}, we assume a Gaussian momentum wave packet $\phi(p)$ of width $\Delta p$ and numerically perform the integrals involved to determine $\mathcal{Q}$ in Eq.~\eqref{eq:Q}, and eventually the output polarization vector and covariance matrix and the phase sensitivity. In this way, we arrive at a prediction for the achievable phase uncertainty $\Delta\varphi$ for a given ensemble size $N$, input squeezing $\xi$, momentum width $\Delta p$ and beam splitter peak Rabi frequency $\Omega_0$.

\subsubsection{Optimal Squeezing Enhancement}

\begin{figure*} 
\begin{center}
\includegraphics[]{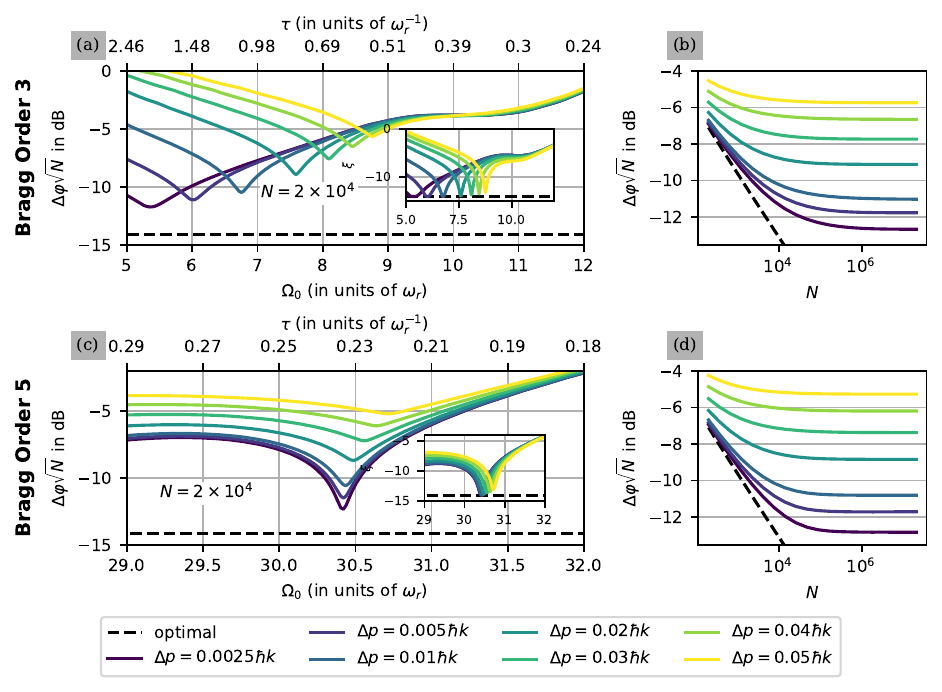}
\end{center}

\caption{\textbf{Simulation of Bragg MZI for Bragg Order 3 and 5} (a), (c) Minimal achievable scaled uncertainty $\Delta \varphi$ obtainable and corresponding optimal squeezing $\xi$ of the input state versus peak Rabi frequency $\Omega_0$ of the two beam splitters in the interferometer sequence, for different momentum widths $\Delta p$ (in units of $\hbar k$) of the input state as indicated in the legend. The number of particles is $N=2\cdot 10^4$. The peak Rabi frequency $\Omega_0$ is given in terms of the recoil frequency $\omega_r=\hbar k^2/2m$ where $m$ is the atomic mass. 
(b) and (d) are showing the optimal uncertainty $\Delta \phi$ over the particle number $N$. The dashed lines represents the maximally achievable gain for a OAT state of the given particle number $N$ in a lossless interferometer. The Bragg orders are 3 for (a), (b) and 5 for (c), (d).}
\label{fig: 4}
\end{figure*}

To quantify the gain beyond the SQL, we will present the scaled phase uncertainty $\Delta\varphi\sqrt{N}$, which would match the squeezing parameter $\xi$ in an ideal, lossless interferometer. However, for an interferometer based on realistic light-pulse operations, some reduction in the entanglement enhancement should be expected. The scaled phase uncertainty $\Delta\varphi\sqrt{N}$ is shown in Fig.~\ref{fig: 4}a,c for Bragg order 3,5 versus the beam splitter peak Rabi frequency $\Omega_0$, for various momentum spreads $\Delta p$. In each case, the level of squeezing $\xi$ has been optimised such as to achieve a maximal sensitivity (minimal $\Delta\varphi\sqrt{N}$). The corresponding optimal input squeezing $\xi$ is shown in the insets for Fig.~\ref{fig: 4}. 

The optima exhibited by Fig.~\ref{fig: 4}a,c with respect to the peak Rabi frequency in the beam splitter operations can be understood from the trade-off inherent to Bragg diffraction in managing the associated losses: Low peak Rabi frequencies require longer pulse durations, which increase Doppler selectivity and lead to significant losses, particularly for wave functions with broad momentum uncertainty $\Delta p$. Conversely, high Rabi frequencies and shorter pulse durations (Raman-Nath regime) result in losses to undesired diffraction orders due to Landau-Zener transitions~\cite{Siem__2020}. Balancing these losses optimally is crucial for maximizing phase sensitivity. 

This balance is important even in an interferometer without correlations~\cite{AdaptedMirror,Szigeti_2012}, but it becomes more critical when using entangled and squeezed states. Unlike a lossless interferometer, where increased squeezing always enhances phase sensitivity, a lossy interferometer achieves optimal performance with a finite level of squeezing, tuned to the specific level of losses. This is because atom losses disrupt quantum correlations, leaving the remaining atoms in a decohered state with increased projection noise, thereby undermining the benefits of squeezing. The insets in Fig.~\ref{fig: 4}a, c show that the optimal level of squeezing indeed depends very sensitively on balancing of losses by a proper control of the beam splitter pulses.

Comparing the input squeezing to the achieved reduction in phase uncertainty, as displayed in Fig.~\ref{fig: 4}a and Fig.~\ref{fig: 4}c, it becomes clear that a significant entanglement enhancement is achieved only for a small momentum spread, that is, low effective temperature of the atomic cloud.
This is further underlined by Fig.~\ref{fig: 4}b and Fig.~\ref{fig: 4}d, where we show the improvements in phase uncertainty in its dependence on the ensemble size $N$, where in each case the Rabi frequency was optimised for lowest uncertainty in the range given by Fig.~\ref{fig: 4}a and b. Largest gains, close to the optimum achievable with OAT squeezed states, will require narrow momentum distributions, with matched levels of squeezing and light-pulse operations.

\begin{figure}
    \centering
    \hspace*{-0.5cm}\includegraphics[]{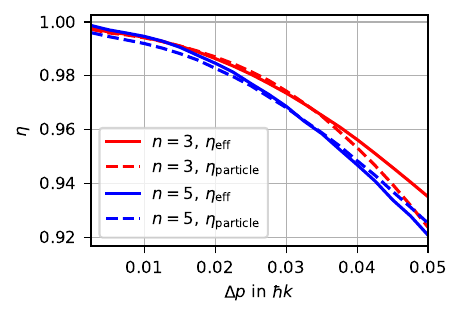}
    \caption{\textbf{Particle Transmission Coefficient $\eta$} We plot the particle transmission coefficients $\eta_\mathrm{eff}$ and $\eta_\mathrm{particle}$ for Bragg order 3 (red) and 5 (blue) by extracting the limit we reach in Figs. \ref{fig: 4} (b) and (d) and using the equations for lossy interferometers in ~\cite{ElusiveHeisenberg,Escher2011} for different momentum widths $\Delta p$ of the input state. The dashed lines are the corresponding particle transmission coefficients $\eta$ calculated by using the relative particle number $\frac{\langle S_0 \rangle}{\langle J_0 \rangle} = \eta_\mathrm{particle}$.}
    \label{fig:etaovern}
\end{figure}

Fig.~\ref{fig: 4}b and Fig.~\ref{fig: 4}d clearly show that in the limit of large particle numbers $N$ certain asymptotic levels of phase uncertainty are achieved, which depend on the initial momentum width $\Delta p$. It is instructive to compare these asymptotic uncertainties to known fundamental bounds for interferometery subject to particle losses. In~\cite{ElusiveHeisenberg,Escher2011} it is shown that the phase uncertainty of an interferometer with an effective particle (or power) transmission coefficient $\eta_\mathrm{eff}$ is bounded by
\begin{align}
\Delta \varphi \ge \sqrt{\frac{1-\eta_\mathrm{eff}}{\eta_\mathrm{eff} N}}.
\end{align}
This bound cannot be directly applied to the atom interferometer considered here, since both loss factors and transmission amplitudes are momentum dependent and therefore not characterized by a single coefficient. Still, it is interesting to use it together with the asymptotics from Fig.~\ref{fig: 4}b and Fig.~\ref{fig: 4}d to infer an effective transmission coefficient $\eta_\mathrm{eff}$. This can then be compared to the particle transmission coefficient $\eta_\mathrm{particle}$ obtained 
by relating the total number of atoms in the output to the one in the input
\begin{align}
\eta_\mathrm{particle} = \frac{\langle S_0 \rangle}{\langle J_0\rangle}.
\end{align}
We use here that the $0$-th components of the output and input pseudospins, defined in Eqs.~\eqref{eq:Sdef} and \eqref{eq:Jdef} respectively, measure total particle numbers. The two values inferred in this way are compared in Fig.~\ref{fig:etaovern} as solid and dashed lines for $\eta_\mathrm{eff}$ and $\eta_\mathrm{particle}$, respectively. There appears to be a decent agreement, despite the fact that the present interferometer falls outside the scope of the fundamental assumptions made in~\cite{ElusiveHeisenberg,Escher2011}.

\section{Conclusion}
In this work, we developed a comprehensive framework for analyzing and optimizing the performance of atom interferometers utilizing OAT spin-squeezed states, specifically within the context of a Mach-Zehnder Bragg interferometer. We derived input-output relations for polarization vectors and covariance matrices, accounting for the non-unitary nature of realistic interferometers. This general formalism actually applies to any interferometer where up to two input ports are populated and two output ports are measured, and thus is relevant well beyond the scenario of a Mach-Zehnder Bragg interferometer considered here as a case study.
In this specific context losses are due to Doppler detuning and undesired diffraction orders, and can be fully accounted for in our theoretical framework. With carefully tuned Gaussian Bragg beam splitters, suitably one-axis twisted spin-squeezed states can significantly improve phase sensitivity. The analysis also makes evident the substantial challenges regarding the effective temperature of the atomic initial state in realizing the full potential of entanglement enhancement. Optimal performance was found to depend on a precise combination of momentum spread, squeezing strength, and light-pulse parameters. These findings underscore the potential of quantum entanglement to advance the sensitivity of atom interferometry, paving the way for more precise measurements in fundamental physics and applied metrology.

As an outlook, we believe that our general input-output formalism provides a suitable basis to design cost functions for systematic optimisations of light pulses beyond the class of Gaussian Bragg pulses, along the lines of~\cite{Szigeti_2012,Louie_2023}, but allowing for entangled input states. Also, the current analysis focuses on squeezed input state, which does not fully exploit the entanglement inherent to OAT states~\cite{RevModPhys.90.035005,Schulte2020}. Extension of the current treatment to states with higher OAT strength along the lines of~\cite{Corgier2021b} would be an attractive possibility. 

\section*{Acknowledgments} 
This project was funded within the QuantERA II Programme that has received funding from the European Union’s Horizon 2020 research and innovation programme under Grant Agreement No 101017733 with funding organisation DFG (project number 499225223 SQUEIS). J.-N. K.-S. and N.G.
acknowledge funding from the EU project CARIOQA-PMP (101081775). J.-N. K.-S. acknowledges funding by the Deutsche Forschungsgemeinschaft (DFG, German Research Foundation) under Germany’s Excellence Strategy – EXC- 2123 QuantumFrontiers – 390837967, through the QuantumFrontiers Entrepreneur Excellence
Programme (QuEEP) and through the CRC 1227 ‘DQ-mat’ within projects A05. KH and NG acknowledge funding by the AGAPES project - grant No 530096754 within the ANR-DFG 2023 Programme. The publication of this article was funded by the Open Access Fund of the Leibniz Universität Hannover.

\appendix

\section{Appendix}

\subsection{Derivation of input-output relation}\label{App1}

Here, we present the derivation of the input-output relations in Eqs.~\eqref{eq:inout}. We start by introducing without loss of generality an orthonormal basis $\{\phi_\xi(p)\}_{\xi\in\mathbb{N}}$ in $L_2(\mathbb{R})$ whose first element $\phi_0(p)$ corresponds to the particular wave function $\phi(p)$ of the input state. We thus have 
\begin{align}
    \sum_\xi \phi_\xi^*(p) \phi_\xi(p') = \delta(p-p').
\end{align}
We can introduce creation and annihilation operators associated to these modes in momentum bin $i$
\begin{align}
    \hat{a}_{i,\xi} = \int dp\: \phi^*_\xi(p) \hat{\psi}_i(p)
\end{align}
where the $\hat{a}_{i,0}$ correspond to the operators introduced in Eq.~\eqref{eq:mode}. Inverting this relation yields
\begin{align}
    \hat{\psi}_i(p) = \sum_\xi \phi_\xi(p) \hat{a}_{i,\xi}
\end{align}

We consider here input states $\ket{\genpsi}$ which only have particles in the first and second port in the mode $\phi_0$, that is
\begin{align}
    \ket{\genpsi} = f(a_{1,0},a_{2,0},a^\dagger_{1,0},a^\dagger_{2,0}) \ket{\mathrm{vac}},
\end{align}
where $f$ is an arbitrary function. For this general class of states we have the convenient property 
\begin{align}\label{eq:psirel}
    \hat{\psi}_i(p) \ket{\genpsi} &= \sum_\xi \phi_\xi(p) \hat{a}_{i,\xi} \ket{\genpsi} \\
    &= \begin{cases} \phi_0(p) \hat{a}_{i,0}\ket{\genpsi}, & \text{for }i\le 2 \\
        0, & \text{else}
    \end{cases}
\end{align}

The complete interferometer sequence is in principle described by a unitary operator $U$ if all (countably infinite) momentum bins $i$ would be taken into account. This $U$ defines a unitary transfer matrix $\Smat(p)$ for the field operators
\begin{align}\label{eq:Urel}
    \hat{U}^\dagger \hat{\psi}_i(p) \hat{U} &= \sum_{j=1} \Smat_{ij}(p) \hat{\psi}_j(p) 
\end{align}
Note that in the main text, we denote by $\Smat(p)$ just the $2\times 2$ sub-block referring to the main interferometer input and output ports. For sake of simplicity, we will use the same symbol here to denote the full (formally infinite dimensional) transfer matrix. What will be shown here, is that only the $2\times 2$ sub-block matters.

The $\alpha$--component of the output polarization vector is
\begin{align}
    {P}_\alpha^\mathrm{out}=\bra{\Psi_\mathrm{out}}\hat{S}_\alpha \ket{\Psi_\mathrm{out}}.
\end{align}
Inserting the definition of the measured pseudo-spin $\hat{S}$ from Eq.~\eqref{eq:Sdef} and using relations \eqref{eq:psirel} and \eqref{eq:Urel} yields
\begin{align}\label{eq: average S_alpha}
     {P}_\alpha^\mathrm{out} 
    &=  \int dp\: \abs{\phi_0(p)}^2 \\
    &\qquad\times\sum_{i,j=1}^2 \sum_{k,l=1}^2 \Smat_{ki}^\dagger (p) \sigma_{ij}^\alpha \Smat_{jl}(p)  \bra{\genpsi} \hat{a}_k^\dagger \hat{a}_l \ket{\genpsi} \nonumber
\end{align}
By introducing the matrix $Q(p)$ via 
\begin{align}\label{eq:Qalphabeta}
     Q_{\alpha \beta}(p) = \frac{1}{2} \sum_{i = 1}^2 \Smat^\dagger_{i j} (p) \sigma^\alpha_{j k} \Smat_{k l}(p) \sigma_{l i}^\beta
\end{align}
we can write
\begin{align}
    \Smat_{ik}^\dagger(p) \sigma^\alpha_{kl} \Smat_{lj}(p) &= \sum_{\beta=0}^3 Q_{\alpha \beta}(p)\sigma^\beta_{ij}. 
\end{align}
The polarization vector component becomes
\begin{align}
    P^\mathrm{out}_\alpha  &= \int dp\: \abs{\phi_0(p)}^2 \sum_{\beta=0}^3 \sum_{k,l=1}^2 Q_{\alpha \beta}(p) \sigma^{\beta}_{kl}  \bra{\genpsi} \hat{a}_k^\dagger \hat{a}_l \ket{\genpsi} \nonumber \\
    &= \sum_{\beta=0}^3 \mathcal{Q}_{\alpha \beta}  \langle \hat{J}_\beta \rangle \label{eq: simp output spin} 
\end{align}
where we used the definition of the pseudo-spin input operator $\hat{J}_\alpha$ in Eq.~\eqref{eq:Jdef} and introduced the matrix $\mathcal{Q}$ via
\begin{align}\label{eq:Qalphabetaint}
    \mathcal{Q}_{\alpha \beta} &= \int dp\: \abs{\phi_0(p)}^2  Q_{\alpha \beta}(p).
\end{align}
With $P^\mathrm{in}_\beta=\langle \hat{J}_\beta \rangle$, the last two equations establish, respectively, Eq.~\eqref{eq:Pinout} and~\eqref{eq:Q} of the main text.

The derivation of the input-output relation for the covariance matrix is somewhat more tedious, but proceeds essentially along the same line. We consider first the non-symmetrized product
\begin{align}
    &\langle \hat{S}_\alpha \hat{S}_\beta \rangle = \bra{\genpsi}\hat{U}^\dagger \hat{S}_\alpha \hat{S}_\beta \hat{U} \ket{\genpsi} \\
     &= \sum_{rstu=1}^2\left( \int dp\: \abs{\phi_0(p)}^2 \sum_{ij=1}^2 \Smat_{ri}^\dagger(p) \sigma_{ij}^\alpha \Smat_{jt} (p)\right) \nonumber\\
     &\qquad\quad\times\left(\int d\Bar{p}\: \abs{\phi_0(\Bar{p})}^2 \sum_{kl=1}^2 \Smat_{sk}^\dagger (\Bar{p}) \sigma_{kl}^\beta \Smat_{lu}(\Bar{p}) \right)\nonumber \\
     &\qquad\quad\times\bra{\genpsi} \hat{a}_r^\dagger  \hat{a}_t \hat{a}_s^\dagger \hat{a}_u \ket{\genpsi} \label{eq: 1. term of 3, 1. split}\\
     &\quad + \sum_{ru=1}^2  \int dp\: \abs{\phi_0(p)}^2 \label{eq: 2. term of 3, 1. split}\\
     &\quad\qquad\times\sum_{ijl=1}^2 \Smat_{ri}^\dagger (p) \sigma_{ij}^\alpha \sigma_{jl}^\beta \Smat_{lu}(p) \bra{\genpsi} \hat{a}_r^\dagger \hat{a}_u \ket{\genpsi}\nonumber\\
     &\quad- \sum_{ru=1}^2 \int dp\: \abs{\phi_0(p)}^2\int d\Bar{p} \: \abs{\phi_0(\Bar{p})}^2\nonumber\\
     &\quad\qquad\times\sum_{ijklt=1}^2 \Smat_{ri}^\dagger (p) \sigma_{ij}^\alpha \Smat_{jt}(p) \Smat_{tk}^\dagger (\Bar{p})  \sigma_{kl}^\beta \Smat_{lu}(\Bar{p})\nonumber\\
     &\quad\qquad\times\bra{\genpsi}\hat{a}_r^\dagger \hat{a}_u \ket{\genpsi} \label{eq: 3. term of 3, 1. split}\\
     &=(\textrm{I})_{\alpha\beta}+(\textrm{II})_{\alpha\beta}-(\textrm{III})_{\alpha\beta}\label{eq:123}
 \end{align}
The last equality invokes again relations \eqref{eq:psirel} and \eqref{eq:Urel} and some algebra. From here on we will focus separately on the three terms in \eqref{eq: 1. term of 3, 1. split}, \eqref{eq: 2. term of 3, 1. split}, and \eqref{eq: 3. term of 3, 1. split}, which we denote by $(\textrm{I})_{\alpha\beta}$,  $(\textrm{II})_{\alpha\beta}$, and $(\textrm{III})_{\alpha\beta}$.

By means of the definition of the matrix $\mathcal{Q}$ in Eqs.~\eqref{eq:Qalphabeta} and \eqref{eq:Qalphabetaint}, and following the logic of the calculation for the polarization vector, one can show
\begin{align}
    (\textrm{I})_{\alpha\beta}
     &= \sum_{\gamma \delta =0}^3 \mathcal{Q}_{\alpha \gamma}   \langle \hat{J}_\gamma \hat{J}_\delta \rangle \mathcal{Q}^T_{\delta \beta}.
\end{align}
Symmetrization with respect to the indices $(\alpha,\beta)$ yields
\begin{align}\label{eq:1sym}
    (\textrm{I})_{\alpha\beta}+(\textrm{I})_{\beta\alpha}&= \sum_{\gamma \delta =0}^3 \mathcal{Q}_{\alpha \gamma}   \left\langle \qty{\hat{J}_\gamma, \hat{J}_\delta} \right\rangle \mathcal{Q}^T_{\delta \beta}.
\end{align}
where $\qty{.,.}$ denotes the anticommutator.

For the other two terms, it will be advantageous to consider directly the symmetrized form and to express the anti-commutator of the Pauli operators by means of the $4\times 4$ matrix $\Lambda(\Vec{\sigma})$ defined componentwise by
\begin{align}
     \left\{\sigma^{\alpha},\sigma^{\beta}\right\} &=  
    2 \qty[\Lambda(\Vec{\sigma})]_{\alpha \beta}.
\end{align}
This implies
\begin{align}
    \Lambda(\Vec{\sigma}) &=  \begin{pmatrix}
        \sigma^0 & \sigma^1 & \sigma^2 & \sigma^3 \\
        \sigma^1 & \sigma^0 & 0 & 0 \\
        \sigma^2 & 0 & \sigma^0 & 0 \\
        \sigma^3 & 0 & 0 & \sigma^0 \end{pmatrix}
\end{align}
With this definition, one can show
\begin{align}
    (\textrm{II})_{\alpha\beta}+(\textrm{II})_{\beta\alpha}&=2\qty[\Lambda\qty(\mathcal{Q}\vec{P}^\mathrm{in})]_{\alpha\beta}\label{eq:2sym}\\
    (\textrm{III})_{\alpha\beta}+(\textrm{III})_{\beta\alpha}&=2\qty[\mathcal{Q}\Lambda\qty(\vec{P}^\mathrm{in})\mathcal{Q}^T]_{\alpha\beta}\label{eq:3sym}
\end{align}

With the definitions for the incoming and outgoing covariance matrices
\begin{align}   
     \Gamma_{\alpha\beta}^\mathrm{in} &= \frac{1}{2} \left\langle \left\{\hat{J}_\alpha, \hat{J}_\beta\right\} \right\rangle - P^\mathrm{in}_\alpha P^\mathrm{in}_\beta
     \\
     \Gamma_{\alpha\beta}^\mathrm{out}&= \frac{1}{2}\left\langle \left\{\hat{S}_\alpha, \hat{S}_\beta\right\} \right\rangle - P^\mathrm{out}_\alpha P^\mathrm{out}_\beta,
\end{align}
and using the symmetrized form of Eq.~\eqref{eq:123} with Eqs.~\eqref{eq:1sym}, \eqref{eq:2sym}, and \eqref{eq:3sym}, we arrive at 
\begin{align}
    \Gamma^{\,\mathrm{out}}_{\alpha\beta} &= \qty[\mathcal{Q} \Gamma^{\,\mathrm{in}} \mathcal{Q}^T + \Lambda (\mathcal{Q}\Vec{P}^{\,\mathrm{in}}) - \mathcal{Q} \Lambda(\Vec{P}^{\,\mathrm{in}}) \mathcal{Q}^T]_{\alpha\beta}
\end{align}
which is Eq.~\eqref{eq:Gammainout}.

\subsection{Interferometer Transfer Matrix $\Smat(p,\varphi)$}\label{App2}
Here, we briefly comment on how the transfer matrix for the interferometer sequence in Eq.~\eqref{eq:transferZ} is constructed. The approach follows closely \cite{Siem__2020,AdaptedMirror}. In order to determine the beam splitter and mirror matrices $\Smat_\mathrm{BS}(p)$ and  $\Smat_\mathrm{M}(p)$ we solve the Schrödinger equation $\dot{U}(t,t_0)=-\i H(t)U(t,t_0)$ with the Bragg Hamiltonian
\begin{align}
    H(t)=\frac{p^2}{2m}+\frac{\hbar\Omega (t)}{2}\qty(\e^{2\i kz}+\e^{-2\i kz})
\end{align}
for a Gaussian pulse 
\begin{align}
    \Omega = \Omega_0 \e^{-\frac{t^2}{2 \tau^2}}.
\end{align}
This is done by expanding on a basis $\qty{\ket{2n\hbar k+p}}$ for $n\in\mathds{N}$ (with a suitable truncation) and $p\in\qty[-\hbar k,\hbar k]$. Due to its periodicity in space, the Hamiltonian couples only states for fixed (quasi)momentum $p$. In this way we infer the relevant transition amplitudes $A_{n,n'}(p)=\mel{2n\hbar k+p}{U(t0,-t0)}{2n'\hbar k+p}$ for a suitable time interval with $t_0\gg\tau$. 
The $2\times 2$ beam splitter and mirror matrices for 3rd-order Bragg diffraction as considerd in the main text are assembled from the elements $A_{3,\pm 3}(p)$ and $A_{\pm 3, 3}(p)$.


%

\end{document}